\documentstyle[twoside,fleqn,espcrc2,psfig]{article}

\newcommand\agt{\hbox{${\lower.40ex\hbox{$>$}
\atop \raise.20ex\hbox{$\sim$}}$}}
\newcommand\alt{\hbox{${\lower.40ex\hbox{$<$}
\atop \raise.20ex\hbox{$\sim$}}$}}

\newcommand\MC{{\,MC}}
\newcommand\PT{{\,PT}}

\title{Perturbative two- and three-loop
coefficients from large $\beta$ Monte Carlo}

\author{G. P. Lepage\address{Newman Laboratory of
Nuclear Studies, Cornell University, Ithaca, NY, 14853},
P. B. Mackenzie\address{Fermilab,
MS 106, P.O. Box 500, Batavia, IL 60510},
N. H. Shakespeare\address{Physics Department,
Simon Fraser University, Burnaby, B.C., Canada V5A 1S6},
and H. D. Trottier$^{\rm c}$}

\begin{document}

\begin{abstract}
Perturbative coefficients for Wilson loops and the static quark
self-energy are extracted from Monte Carlo simulations at large
$\beta$ on finite volumes, where all the lattice momenta are large.
The Monte Carlo results are in excellent agreement with perturbation
theory through second order. New results for third order coefficients
are reported. Twisted boundary conditions are used to eliminate
zero modes and to suppress $Z_3$ tunneling.
\end{abstract}

\maketitle

\section{INTRODUCTION}

Simulations using highly improved lattice actions have become
commonplace in recent years. Higher order perturbative
calculations for these actions present a major challenge
however, which must be overcome in order to obtain
precision results for a number of observables.

A simple alternative to analytic perturbation theory,
proposed in Ref.\ \cite{Dimm}, is to directly measure
observables in Monte Carlo simulations at
very large $\beta$ 
and to fit the coefficients of the perturbative expansion to the results.
This produces coefficients with both statistical and truncation
errors, so it is not a complete substitute for conventional
perturbation theory.  However, as we show here, it
produces estimates of high order coefficients with far less
effort than a conventional calculation.
This method was shown to reproduce
analytical results for the one-loop mass renormalization
for Wilson fermions, and the one-loop additive energy for
NRQCD fermions \cite{Dimm}. In addition, preliminary estimates
of some third-order Wilson loop coefficients have been made
\cite{DimmWilsonLoops}. An extension of this technique to
background field calculations has also been considered
\cite{TrottierLepage}.

In this work we make a much more extensive analysis of
this method. We reproduce known second-order perturbative
coefficients for Wilson loops and the static quark self-energy to
high accuracy. New predictions for third-order coefficients for
these observables are made, and results of extensive systematic
studies are reported.

\section{WILSON LOOPS}

To begin with, we present results of simulations of Wilson loops
for the Wilson gauge-field action on $16^4$ lattices.
Periodic boundary conditions were used
here in order to make a direct comparison with the second order
perturbative coefficients calculated by Heller
and Karsch \cite{Heller} on the same lattice volume.
Simulations were done at nine couplings, from $\beta \approx 9$
to $\beta \approx 50$.

We analyze the logarithm of the $R\times T$ Wilson loop
\[
- {1\over 2 (R+T)} \ln W_{R,T} = \sum_n c_n \,
\alpha_P^n (q^*_{R,T}),
\]
scaled such that the coefficients for large loops
approach those of the static self-energy.
We use a renormalized coupling $\alpha_P$ determined from
measured values of the plaquette \cite{LepMac,NRQCDalphas}
\[
-\ln W_{1,1}^\MC \equiv
{4\pi \over 3} \alpha_P(3.41/a)
\left[ 1 - 1.185 \, \alpha_P \right] .
\]
This eliminates large renormalizations of the bare
lattice coupling $\alpha_0 = 3 / (2\pi \beta)$, as can be seen
by relating $\alpha_P$ to $\alpha_0$, using the third-order
expansion of the plaquette given in Ref. \cite{Alles}:
\begin{equation}
   \alpha_P(3.41/a) \approx \alpha_0 + 4.558 \, \alpha_0^2
                              + 28.499 \, \alpha_0^3 .
\label{eq:aPtoa0}
\end{equation}
The renormalized coupling is evaluated at scales $q^*_{R,T}$
determined by the procedure of Ref. \cite{LepMac}.
It is also possible to perform the fitting procedure using the bare
coupling constant, $\alpha_0$, but this leads to much larger
higher order coefficients and truncation errors because of the 
large coefficients in Eqn.~\ref{eq:aPtoa0}.
These lead in turn to poorer fits with larger $\chi^2$'s.

\begin{figure}[htb]
% \vspace{-9pt}
\psfig{file=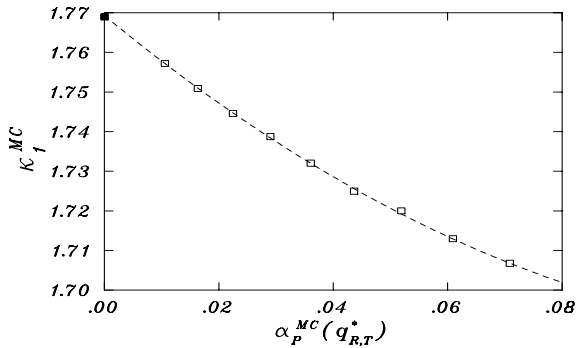,height=4.5cm,width=7.5cm,angle=90}
%%\vspace{-24pt}
\caption{Monte Carlo results for $\kappa_1$ for the $5\times5$ Wilson loop,
with best fit.}
% \vspace{-9pt}
\label{fig:ProcWilsonK1}
\end{figure}

It is very straightforward to fit the leading perturbative
coefficients to the Monte Carlo data.
Monte Carlo results for the $5\times5$ Wilson loop are presented
in Fig.\ \ref{fig:ProcWilsonK1}, in terms of the quantity
\[
\kappa_1 \equiv
 - \ln W_{R,T}^\MC / \left( 2 (R+T) \alpha_P(q^*_{R,T}) \right) ,
\]
after the known first-order contribution
due to zero modes \cite{Coste} is subtracted from the data.
The dashed line in Fig.\ \ref{fig:ProcWilsonK1} is the
result of a fit to
$\kappa_1 \approx c_1 + c_2 \alpha_P + c_3 \alpha_P^2$;
the filled square shows the analytical value of $c_1$, while
the curvature in the results for $\kappa_1$ demonstrates
a signal for $c_3$.

The measured values of $c_{1,2}$ for various small loops are
shown in Table \ref{table:Wilsonc12}.  They
are in excellent agreement with perturbation theory.
\begin{table*}[hbt]
%%\vspace{-9pt}
\caption{Perturbative coefficients $c_{1,2}$ of Wilson loops from 
 Monte Carlo simulations and perturbation theory.}
\label{table:Wilsonc12}
\begin{center}
\begin{tabular}{|c||c|c||c|c||c|}
\hline
     Loop &  $c_1^\MC$    &  $c_1^\PT(q^*_{R,T})$
          &  $c_2^\MC$    &  $c_2^\PT(q^*_{R,T})$  & $q^*_{R,T}$ \\
\hline
  $1\times2$  &   1.2037(1) &  1.2039  &  -1.248(08) & -1.257 &3.07\\
  $1\times3$  &   1.2586(2) &  1.2589  &  -1.186(09) & -1.195 &3.01\\
  $2\times2$  &   1.4336(2) &  1.4338  &  -1.314(08) & -1.320 &2.65\\
  $3\times3$  &   1.6088(3) &  1.6089  &  -1.205(12) & -1.204 &2.46\\
  $4\times4$  &   1.7065(4) &  1.7066  &  -1.202(16) & -1.198 &2.30\\
  $5\times5$  &   1.7692(6) &  1.7690  &  -1.191(36) & -1.166 &2.23\\
\hline
\end{tabular}
\end{center}
%%\vspace{-18pt}
\end{table*}

\begin{figure}[htb]
%%\vspace{-18pt}
\psfig{file=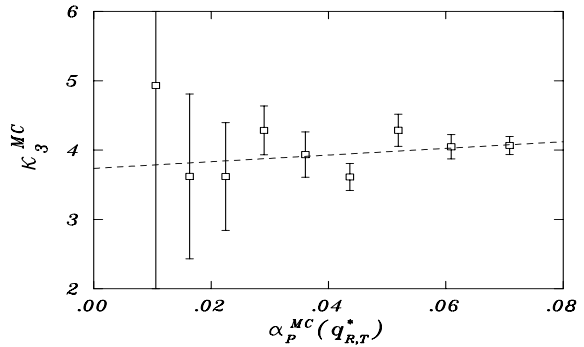,height=4.5cm,width=7.5cm,angle=90}
%%\vspace{-18pt}
\caption{Monte Carlo results for $\kappa_3$ for the $5\times5$ Wilson loop,
with best fit.}
%%\vspace{-15pt}
\label{fig:ProcWilsonK3}
\end{figure}

We also make new predictions for the third-order
coefficients $c_3$. To improve the accuracy of these results,
it is helpful to subtract the analytically known first- and second-order
perturbative contributions from the data.
Monte Carlo results for the residual
\[
\kappa_3 \equiv {1 \over \alpha_P^3}
\left[ - {1\over 2 (R+T)} \ln W_{R,T}^\MC
- c_1 \alpha_P - c_2 \alpha_P^2 \right]
\]
are shown in Fig.\ \ref{fig:ProcWilsonK3} for the
$5\times5$ loop, with the results of a fit to
$\kappa_3 \approx c_3 + c_4 \alpha_P$.

The data are not accurate enough to resolve $c_4$,
although the best fit errors indicate that $c_4$ is of
the same order as $c_3$ for these Wilson loops.
To obtain errors on the $c_3$ that take into account the
effects of truncation in $\alpha_s$, we include the $c_4$ term
in our fits. In
the results for $c_3$  given in Table \ref{table:Wilsonc3},
the quoted errors come from the fit including $c_4$, which
effectively includes an uncertainty due to the possible
values of $c_4$ allowed by the data.
\begin{table}[hbt]
%%\vspace{-18pt}
\caption{Monte Carlo results for third order coefficients $c_3$
of Wilson loops.}
\label{table:Wilsonc3}
\begin{center}
\begin{tabular}{|c|c|c|}
\hline
    Loop  &  $c_3^\MC(q^*_{R,T})$ & $q^*_{R,T}$\\
\hline
  $1\times2$  &   0.40(08)   &	3.07  \\
  $1\times3$  &   0.63(10)   &	3.01  \\
  $1\times4$  &   0.82(10)   &	2.97  \\
  $1\times5$  &   0.92(10)   &  2.95  \\
  $2\times2$  &   1.38(09)   &	2.65  \\
  $2\times3$  &   1.89(11)   &	2.56  \\
  $2\times4$  &   2.26(11)   &	2.49  \\
  $2\times5$  &   2.51(11)   & 	2.46  \\
  $3\times3$  &   2.49(12)   &	2.46  \\
  $3\times4$  &   2.94(11)    &	2.38  \\
  $3\times5$  &   3.23(13)    &	2.35  \\
  $4\times4$  &   3.35(14)    &	2.30  \\
  $4\times5$  &   3.67(19)    &	2.27  \\
  $5\times5$  &   3.74(33)    &	2.23  \\
\hline
\end{tabular}
\end{center}
%%\vspace{-18pt}
\end{table}

We used periodic boundary conditions in these tests on
small Wilson loops in order to be able to compare directly
with Ref.~\cite{Heller}.
The known leading effects of the resulting zero modes
were subtracted by hand.   We have not
corrected for the contribution from zero modes beyond first order;
these effects should be very small for small Wilson loops.
In fact there is no visible effect of zero
modes beyond first order, within statistical errors. (This
would show up in singular behavior in
$\kappa_3$ at small $\alpha_P$).

\section{STATIC QUARK SELF-ENERGY}

To calculate quantities more complicated than Wilson loops,
we need to solve some additional problems.
In general, the effects of zero modes are unknown and not
small, so we must remove them from the first.
Further, most quantities are highly sensitive to the effects of
tunneling between the various $Z_3$ vacua of $SU(3)$,
so we must suppress this, too.  We test our ability to
do this by calculating the static quark self-energy $E_0$.

Zero modes can be eliminated by using twisted boundary
conditions \cite{Luscher}, and remaining finite size effects
can be removed by extrapolating results from several lattice
volumes. 

We measured the gauge-invariant
Polykaov line on lattices of various volumes $L^3 \times (T = L)$
\[
   P_4(L) \equiv {1 \over 3 L^3} \sum_{\vec x}
            \mbox{ReTr} \prod_{x_4=1}^L U_4(x) .
\]
Defining $E_0(L) \equiv -\ln P_4(L) / L$,
we have $E_0 \equiv E_0(L\to\infty)$.
The two-loop expression for $E_0$ can be obtained
from perturbative results for $P_4$ \cite{Heller,Martinelli}
\[
   E_0 = 2.1173 \, \alpha_P(1.68/a)
       - 1.124 \,  \alpha_P^2  +  O(\alpha_P^3) .
\]

The Polyakov line is an order parameter for the $Z_3$
degenerate vacua of the gauge theory, and has different
values in  different $Z_3$ phases.
To make its vacuum expectation value well defined 
and nonzero in a Monte Carlo simulation,
one can add an external field to the action to
pin the simulation into one of the $Z_3$ states
and take the limit of the external field going to zero
as the volume goes to infinity.

In order to minimize these nonperturbative effects,
we start the simulation with all links
initialized to $U_\mu = I$; tunneling and domain
formation are suppressed by working at sufficiently large
lattice volumes and couplings $\beta$. 
We found that with periodic boundary conditions, 
tunnelings were quite frequent, even at the surprisingly large 
$\beta$'s of 50 -- 100.
On the other hand, we find that
twisted boundary conditions lead to a dramatic suppression
of these effects.
Twisting in two directions strongly suppresses tunneling
and twisting in three directions
virtually eliminates it, even on runs of hundreds of thousands
of sweeps and relatively low $\beta$.
Thus, so far, adopting the pinning method has been unnecessary.

\begin{figure}[htb]
%%\vspace{-9pt}
\psfig{file=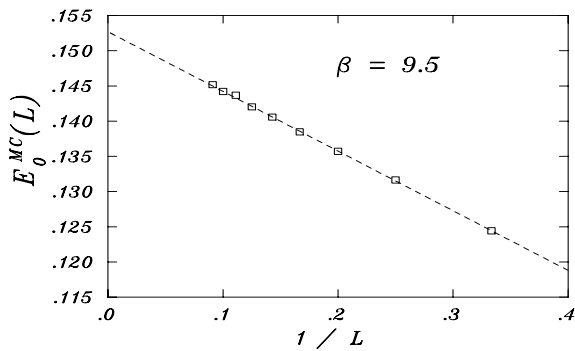,height=4.5cm,width=7.5cm,angle=90}
%%\vspace{-18pt}
\caption{Infinite volume extrapolation of the static quark self-energy
 $E_0(L)$.}
%%\vspace{-15pt}
\label{fig:ProcE0L}
\end{figure}

Finite-size effects are consistent
with perturbative expectations. We estimate $E_0$ by fitting $E_0(L)$
to the perturbative form of the interaction between the static quark
and its images in the box walls:
\[
   E_0(L) = E_0 - (a_1 + a_2\ln(L)) / L .
\]
Simulations were done with twisted boundary conditions
on 9 volumes, $L=[3,11]$. Measurements were made
at 9 couplings on each volume, from $\beta \approx 9$ to
$\beta \approx 60$. Results for $E_0(L)$ at $\beta=9.5$
are shown in Fig.\ \ref{fig:ProcE0L}; in all cases, the fits yield
$a_1 = O(\alpha_P)$ and $a_2 = O(\alpha_P^2)$, as expected.

\begin{figure}[htb]
%%%\vspace{-9pt}
\psfig{file=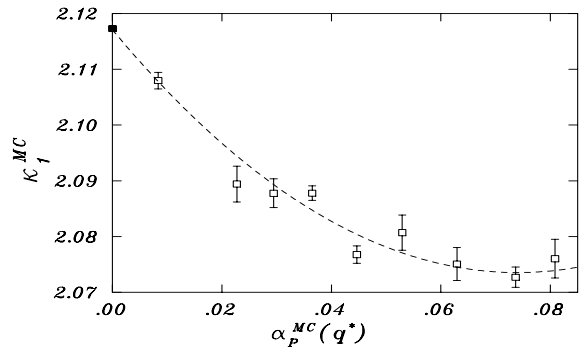,height=4.5cm,width=7.5cm,angle=90}
%%\vspace{-24pt}
\caption{Monte Carlo results for $\kappa_1$ for the static quark self-energy,
with best fit.}
%%\vspace{-18pt}
\label{fig:ProcE0K1}
\end{figure}

Results for $\kappa_1 \equiv E_0 / \alpha_P(q^*=1.68)$
are shown in Fig.\ \ref{fig:ProcE0K1}. Best fits to
$\kappa_1 \approx c_1 + c_2 \alpha_P + c_3 \alpha_P^2$
yield $c_1^\MC = 2.117(3)$ and $c_2^\MC = -1.18(18)$, in good
agreement with perturbation theory.
We also obtain a new prediction for the third order
term; using the analytical values for $c_1$ and $c_2$, the
best fit yields 
\begin{equation}
c_3^\MC = 5.6(1.8).
\label{eq:E0}
\end{equation}
As for the loops, we include a fourth order term in the fit.
This guards against the presence of an anomalously large $c_4$,
and enlarges the error bar.

A result for the third order term in the expansion of
$E_0$ in the bare lattice coupling has also been reported
in this conference, using coupled Langevin equations
\cite{Parma}. Large coefficients in the expansion in
terms of $\alpha_0$ reflect the need to renormalize
the coupling; for example, combining our result for
$c_3^\MC$ with Eqn. (\ref{eq:aPtoa0}), we find
\begin{equation}
E_0 \approx 2.1173 \, \alpha_0 + 11.152 \, \alpha_0^2
+ 81(2) \, \alpha_0^3.
\label{eq:E00}
\end{equation}
The absolute errors in Eqns.~\ref{eq:E0} and \ref{eq:E00}
are the same.
The relative error in Eqn.~\ref{eq:E00} seems surprisingly small
because the size of the coefficient itself has been 
increased by analytic terms from Eqn.~\ref{eq:aPtoa0}.

\section{ALGORITHMIC INVESTIGATIONS}

These calculations have been done using standard pseudo-heat bath
algorithms to create the gauge fields.
The calculation of $E_0$ took a little over a month running
on about 20 PC's.
Clearly finding the best simulation algorithms is an essential
element of the method.
In this section, we report some preliminary investigations into
algorithms specifically appropriate to perturbative simulations.

In principle, microcanonical based methods \cite{Cal82} have the potential
to outperform ordinary heat bath methods since they have a critical
scaling dimension of 1 instead of 2.
In practice, for QCD they don't achieve this.
Correlations between nearly degenerate modes last a long time,
and nonperturbative effects cause averages over uncorrelated configurations
to add up statistically, as $1/\sqrt{n}$.
If we examine the perturbative case carefully, however, we see
that this need not be the case.

A single {harmonic oscillator} evolving under a microcanonical
algorithm approaches its average as $1/n$:
\begin{equation}{
\label{eq:osc}
\langle \phi \rangle =  \frac{1}{n}\sum_{m=0}^n\exp(i m \delta)
  = \frac{1}{n}\frac{1-\exp(i n \delta)}{1-\exp(i  \delta)}.
}\end{equation}
Therefore, {free field theory}, which is just a collection of
independent oscillators, does the same.
However, the usual microcanonical evolution equations,
\begin{equation}{
H\equiv T+S,  \ \ \ \ \ T\equiv\frac{1}{2}\sum_i p_i^2,
\ \ \  \ \ S\equiv S(\phi_i)
}\end{equation}
{$\rightarrow$
\begin{equation}{
\frac{dp_i}{d\tau} = -\frac{\partial S(\phi)}{\partial \phi_i},
\ \ \ \ \ \ \frac{d\phi_i}{d\tau}=p_i,
}\end{equation}
do not make a good simulation algorithm for free field theory.
Many sets of momentum modes, e.g., $(1,0,0,0)$ and $(0,1,0,0)$, 
have same the frequencies and never decorrelate,
so that configuration space never gets completely covered.

Since the ``momenta'' in microcanonical updating of
quantum field theories are fictitious,
this problem and be removed by introducing
a ``nondegenerate microcanonical method'' using
 randomized fictitious masses ${m_i}$:
\begin{equation}{
H\equiv T+S,  \ \ \ \ \ T\equiv\frac{1}{2}\sum_i \frac{p_i^2}{{m_i}},
\ \ \ \ \ S\equiv S(\phi_i),
}\end{equation}
{$\rightarrow$ 
\begin{equation}{
\frac{dp_i}{d\tau} = -\frac{\partial S(\phi)}{\partial \phi_i},
\ \ \ \ \ \ \frac{d\phi_i}{d\tau}=\frac{p_i}{{m_i}},
}\end{equation}
 $a_1 < {m_i} \le a_2,$ where $a_1$ and $a_2$ are of order 1,
say 0.8 and 1.2.
Mathematical classical mechanics often starts from periodic system 
system with nondegenerate  frequencies.
Such systems have more tractable mathematics,
for example no zeros in perturbation theory denominators.
With these equations, the
frequencies of free field theory are 
nondegenerate and configuration space is covered densely.
The {KAM theorem} then tells us that
 for {small perturbations} of nondegenerate
unperturbed system, dense tori covered by system's evolution
through phase space deformed only slightly.
Therefore we expect the covering of configuration space to remain
dense in weakly coupled systems, 
and the convergence to the average to continue to go
as $1/n$.
It is easy to test that for { scalar field theory},
 things work out in more or less this way.

In QCD, however, gauge symmetry creates a new complication.
The random mass method may be adapted to gauge theories in two ways
It is straightforward to implement gauge covariant random masses that break
most but not all mode degeneracy:
\begin{equation}{
 T\equiv\frac{1}{2}\sum_i \frac{Tr H_i^2}{{m_i}},
 \ \ \ \ \ S\equiv S(U)
}\end{equation}
$\rightarrow$ 
\begin{equation}{
i\frac{dH_i}{d\tau} = -\frac{\partial S(U)}{\partial U_i},
 \ \ \ \ \frac{dU_i}{d\tau}=\frac{1}{{m_i}} i H U,
}\end{equation}
 $H_i = \sum_a \lambda_a h^a_i$.

Mode degeneracy may be broken completely by
 color breaking random masses:
\begin{equation}{
 T\equiv\sum_{{ a}i} 
   \frac{ {h^{ a}_i}^2}{{m_i^{a}}},
 \ \ \ \ \ S\equiv S(U)
}\end{equation}
$\rightarrow$ 
\begin{equation}{
i\frac{dH_i}{d\tau} = -\frac{\partial S(U)}{\partial U_i},
\ \ \ \ \frac{dU_i}{d\tau}= 
i \sum_a \frac{h^{ a}_i \lambda^a}{{m_i^{a}}} U.
}\end{equation}
However, they  also
break gauge invariance.  This necessitates gauge fixing, 
which is more work and may introduce extra correlations.
\begin{figure}[htb]
%%\vspace{-18pt}
\psfig{file=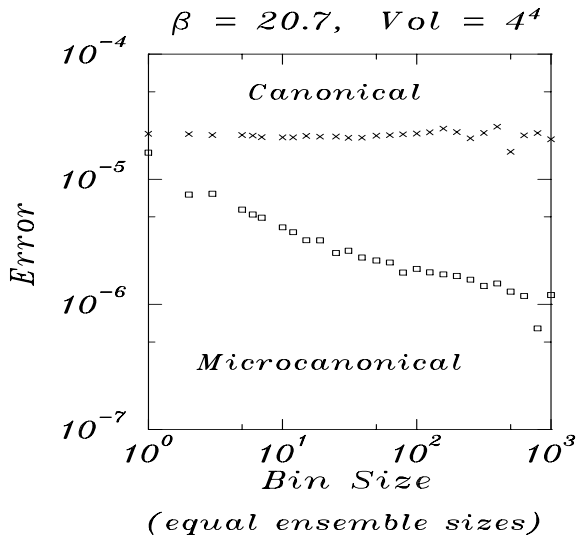,height=7cm,width=7.5cm,angle=0}
%%\vspace{-18pt}
\caption{Statistical error on the plaquette obtained with standard
binning procedure from the heat bath algorithm (canonical) 
and the nondegenerate microcanonical algorithm (microcanonical).
}
%%\vspace{-15pt}
\label{fig:CanAndMicro}
\end{figure}

We have investigated the first of these methods.
New complications are  present in gauge theories.
Free field behavior can only be approached as $\beta\rightarrow\infty$,
$U_P\rightarrow 1.$
But at very high $\beta$, round off becomes important, 
which destroys the anticorrelations leading to $1/n$ behavior.
At small $\beta$, $Z_3$ tunneling occurs.  Nonperturbative effects
can (and do) destroy the anticorrelations,
restoring $1/\sqrt{n}$ statistical behavior.
Hence, twisted periodic boundary conditions are required.
At moderate $\beta$, we find at least partial success with the plaquette.
Fig.~\ref{fig:CanAndMicro} shows the error obtained on the plaquette
using standard binning procedures from the heat bath algorithm 
(labeled canonical) and the nondegenerate microcanonical algorithm
(labeled microcanonical).  When anticorrelations as in Eqn.~\ref{eq:osc}
are present, standard formulae, which assume $1/\sqrt{n}$ behavior,
should produce errors that drop as $1/\sqrt{n}$ in bin size, 
as observed.
The results obtained are $W_{1,1}=0.900346(1)$ with the microcanonical
method and 0.900271(23) with the canonical method for
equal ensemble sizes.
We have not extrapolated to infinite volume, which we believe
accounts for the small discrepancy.

The plaquette is speeded up beautifully, 
achieving the full $1/n$ approach to its average
even with gauge covariant random masses.
However, the plaquette turns out to be a special case,
presumably because it is also the action of the theory.
For other loops, the great speed up is not observed,
which is disappointing but not entirely unexpected
in view of the remaining correlations.
Full breaking of mode degeneracy appears necessary for most quantities.
We are encouraged by the results with the plaquette, but
more work is necessary to make this a general method.

\section{SUMMARY}

In summary, results presented here demonstrate that
second order perturbative coefficients
are readily accessible in Monte Carlo simulations at
large $\beta$. Reasonable estimates of third-order terms
can also be made, especially when analytic results for
lower order terms are available.
We are currently working on gauge-fixed quark
propagators, in order to measure second order
mass renormalizations for NRQCD and Fermilab fermions.

We thank Urs Heller for generously providing us
with his programs for second-order coefficients.
This work was supported in part by the
National Science Foundation, the Department of Energy, and the
National Science and Engineering Research
Council of Canada.

\end{document}